# The 4+1 Model of Data Science

Rafael C. Alvarado
School of Data Science
University of Virginia

11 November 2023

**Introduction**

Data Science is a complex and evolving field, but most agree that it can be defined as a combination of expertise drawn from three broad areas—computer science and technology, math and statistics, and domain knowledge—with the purpose of extracting knowledge and value from data. Many also associate it with a series of practical activities ranging from the cleaning and "wrangling" of data, to its analysis and use to infer models, to the visual and rhetorical representation of results to stakeholders and decision-makers.

This essay proposes a model of data science that goes beyond the laundry-list definitions that dominate the discourse on the topic today. Although these are not inaccurate per se, they do not get at the specific nature of data science or help distinguish it from adjacent fields such as computer science and statistics—fields whose members sometimes claim to already be doing data science. Without a clear understanding of its specific and unique nature, the field is subject to counterproductive turf battles within the academy, confusion in the workplace, and miscommunication between the two sectors. A good definition will increase the alignment between what hiring managers in industry understand the role of data scientist to be and the definitions, often unarticulated, that undergird the academic programs that produce data scientists.

We define data science as an interdisciplinary field comprising four broad areas of expertise. These are the areas of **value**, **design**, **systems**, and **analytics**. A fifth area, **practice**, integrates the other four in specific contexts. We call this the 4+1 model of data science. Together, these areas belong to every data science project, even if they are often unconnected and siloed in the academy. In addition to forming an interdisciplinary model among themselves, each area of the field is internally interdisciplinary as well, bringing together diverse and sometimes contrary perspectives under a common heading. The inherently interdisciplinary and pluralist nature of these areas is a distinctive feature of data science and a key differentiator between it and traditional disciplines. This essay describes how this model is derived and provides clues about how to interpret and apply the model to your own situation.

**A Common Motif**

A review of the literature on data science definitions, including sources from the adjacent fields data analysis and data mining, reveals that most invoke the motif of a data processing *pipeline*—a sequence of events through which data flows as it moves from the consumption of so-called raw data to production of useful results. In this view, consumed data may come from a variety of sources, such as databases or intentional experiments or sensors, and results may be equally various, from the visual communication of analytical results to stake-holders to the development of an interactive data product for use on the web. A detailed review of these sources may be found in the Appendix.

For analytic purposes, it is helpful to view the pipeline motif as a kind of story, a micronarrative that encodes and socializes an understanding about the practice of data science. All forms of work involve the sharing of stories in one form or another, both formally in the context of education and informally through the countless acts of training, mentorship, and imitation that go on everyday in the academy and the workplace. When we want to explain to a trainee, a peer, or a supervisor how something is done, we often resort to a story that is general enough to apply to a variety of contexts yet specific enough to be translatable into action. The data processing pipeline is like that, and the essays in which it appears are part of the ongoing discourse by which a community of practice grows and learns.

By thinking of the pipeline as a story we can analyze it with techniques developed by sociolinguists, folklorists, and students of literature to study how stories are structured and how they function socially and cognitively. One idea from this field is that stories can be broken down into elementary units, or "event functions," and that these functions tend to have an invariant order within specific traditions (Propp and Dundes 1977). In addition, a given sequence of functions may exhibit ab overall patterns, such as a chiasmus structure, in which the functions at the beginning of a story mirror those a the end (Lévi-Strauss 1955). The following analysis applies these ideas.

**The Primary Standard Sequence**

An analysis of a representative sample of essays that define a data processing pipeline shows that the various pipeline stories consist of elements drawn from a primary standard sequence of about twelve elements, give or take a few depending on how one might expand or contract terms. These are listed and defined below by a core set of event types, represented as verbs, with the understanding that many synonyms are employed in the examples.

1. Understand. The work of developing a question that the acquisition and analysis of data can answer. Questions may range from the status of a scientific hypothesis the merit of a business value proposition. More broadly, this is the stage of framing and scoping the work that follows.
2. Plan. The work of designing an experiment to produce data or of devising a strategy to acquire existing data relevant to the goals specified by the Understand phase.
3. Collect. The work of producing or acquiring data. This covers a broad range of activities from taking surveys to generating signal data to harvesting data from databases to scraping websites.
4. Store. The work of moving data from its collected state to a persistent store to house the data. This may be a file system or a database.
5. Clean. The work of fixing problems with data as collected and stored. This covers a wide range of activities, including the normalizing the data formats (such as dates) to the handling of missing data to the transformation of tables into some level of normal form.
6. Explore. The work of investigating cleaned data for general patterns, distributions, simple clusters, outliers, sanity, etc. Exploratory data analysis provides a useful set of tools and perspectives for this work. In some models, this activity is expanded into more sophisticated forms of pattern discovery and becomes associated more directly with the Model phase. Visualization using tools such as box plots, scatter plots, histograms, etc. play a large role here.
7. Prepare. The work of transforming data into a form suitable for the particular methods planned in the Model phase. This may mean generating tidy tables for statistical analysis, wide-format data for matrix operations, one-hot encoded feature spaces for input into deep learning systems, or graph structured data for network analysis. Also included here is feature engineering and dimensionality reduction.

8. <u>Model</u>. The work of using data to train models or estimate parameters for the tasks of prediction, inference, etc. Internally, this phase comprises a sub-pipeline that operates closely with the previous and following steps and involves model selection, parameter tuning, validation, and testing. Additionally, this area is characterized by general approach, which may be based in classification (predictive modeling), inference, or simulation.

9. <u>Interpret</u>. The work of making sense of the results of modeling the data, in terms of the terms set out by the Understand phase. This may mean making claims about causality, assigning real-world correlates to discovered patterns, or assessing the generalizability of a classifier.

10. <u>Communicate</u>. The work of representing data, model, and results to stakeholders as defined or implied by the Understand phase. The includes the work of visualization broadly conceived, from the creation on interactive visualizations to static infographics. It also includes story-telling and essay writing for both specialized and generalized audiences.

11. <u>Deploy</u>. If the work results in a data product, such as real-time classifier to support a business process, this means standing up and maintaining the software product in a performant and scalable hardware environment.

12. <u>Reflect</u>. Between iterations of the pipeline, this is the work of reflecting on the strategic or philosophical significance of the work conducted. It also includes the ethical dimension of the work conducted.

Note that these descriptions are not meant to be prescriptive; they are summaries of the descriptions found in the literature. With this composite view in place, it will be easier to address limitations and fill gaps in our effort to create an authentic definition of data science.

No one definition includes them all, but some are more comprehensive than others, and different disciplines emphasize different parts.

For example, Hayashi's statistically-oriented definition of data science includes just three phases—design [1], collect [3], and analyze [8,9]—with an emphasis on the experimental design phase in which data are actually produced through thoughtfully designed experiments (Hayashi 1998).

Mason and Wiggins propose five—obtain [3,4], scrub [5], explore [6], model [8], and interpret [9]—which highlights two conditions that define industrial data science, the simultaneous availability of data (one obtains it, say through web scraping) and their poor condition relative to analysis, i.e. the need to scrub and wrangle them into usable form (Mason and Wiggins 2010b).

The CRISP-DM model is the most comprehensive, with seven phases defined (if we include the unnamed but visually depicted function of storage), emphasizing the importance of understanding both the business proposition and data before anything is done with it (Wirth and Hipp 1999a). It also modifies the metaphor of the pipeline, representing it as a circular and iterative process. However, unlike Donoho's similarly comprehensive sequence (implied by the ordering of his six divisions of "greater data science"), it does not include a "meta" phase devoted to reflecting on the process as a whole (Donoho 2017b).

As the CRISP-DM model shows, in some cases the pipeline is described as a circular process, where the output of one cycle serves as input to the next. Nevertheless, in these cases the sequential order of processing stages is maintained. Moreover, the image of the circle makes explicit what is implicit in the non-circular models, namely that the pipeline is never a one-off operation but repeated many times in the life-cycle of a project. What the circular models uniquely introduce is the idea that data processing is often a cybernetic process, a feedback loop where the results of one cycle inform the next. However, it should be noted that the idea of feedback foregrounded by circular models is not well developed. We shall see that the relationship between the endpoints is an area deserving of further research and theoretical development.

**Seven Chapters**

A close look at the twelve phases listed above shows that a given phase may be more closely related to some phases than to others—in other words, the phases can be grouped thematically. For example, it seems clear that the Understand and Plan phases go together, just as Clean, Explore, and Prepare do. They belong together because we can imagine performing their associated activities together, and separately from the other phases. Also, in some cases the order of the phases within the group may change—for example, explore might precede cleaning in the event, as the two are often performed simultaneously—while the sequential order of the groups is less likely to vary. Finally, we can imagine assigning different teams to perform the labor in each thematic grouping, by virtue of the expertise required to carry them out.

Given this, the twelve-part composite pipeline can be reduced to seven groups:

*A* understand and plan
*B* collect and store
*C* clean, explore, and prepare
*D* model and interpret
*E* communicate
*F* deploy
*G* reflect

Each of these thematic groups may be considered a "chapter" in the story. Note that the number of verbs in each chapter title does not necessarily predict the length of its content. For example, the chapter on "model and interpret" covers a wide range of activities from a variety of perspectives, including classical statistics, machine learning, and computational simulation. It's a big and complicated chapter with its own pipeline, but it is just one chapter among seven.

**An Arc with Four Zones**

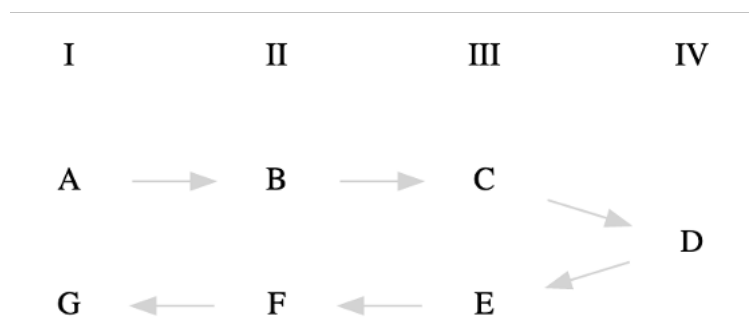

Figure 1. The Standard Sequence as a Narrative Arc

To be sure, the middle chapter plays a central role in our story. If we think of the story as following a classical "there and back again" structure—a chiasmus pattern like $X_1, Y_1, Z, Y_2, X_2$—then chapter $D$ is the pivot, while chapters $A, B,$ and $C$ mirror $E, F,$ and $G$. Thinking of the story in this way allows us to identify a parallel structure in the pipeline, connecting phases that are usually seen as separate. Specifically, we may visualize the pipeline as an arc, a **U**-shaped process in which chapters in the first half of the pipeline mirror the those of the second half. We may then group chapters by the pairs formed in this way, yielding four zones—$A$ and $G$ belong to zone $I$, $B$ and $F$ to $II$, $C$ and $E$ to $III$, and $D$ to $IV$—as in the following diagram:

| | | |
|---|---|---|
| I | understand, plan | reflect |
| II | collect, store | deploy |
| III | clean, prepare, explore | communicate |
| IV | model, interpret | |

Figure 2. The Arc Transposed

With this visualization, we can discern some interesting properties about the data science pipeline that are not obvious in the original sequential image. For one, the arc structure suggests that the two ends of the pipe are not separate; both make direct contact with the external world (relative to the internal world of the pipeline considered as a system). The external world—natural or social—from which data are pulled is the same world into which data products are inserted. This insight echoes the CRISP-DM model, which connects $A$ and $G$ (actually $F$), except that the two ends of the arc model are not directly connected. Instead, they come into contact with—and are separated by—the world in all of its complexity and unpredictability. The relationship between the effects caused by our data products $G$ and the data we pull from the world $A$ is not given but a matter of discovery—and often surprise. This is an important distinction between the model being proposed here and the simple idea that the pipeline forms a closed circle.

**Interpreting the Themes**

At this point, we can explore the unifying themes associated with the four zones in our arc model by transposing the preceding visualization, which draws attention to what is common to each pairing. This generates four candidate areas of data science expertise—activities that, although they appear on opposite ends of the pipeline, nevertheless share basic knowledge, know-how, and areas of concern.

Zone $IV$ is the easiest to interpret in this way because as the pivot of the arc it is not paired. It represents the work of modeling a problem mathematically, as well as evaluating and interpreting the results of mathematical modeling. This work requires data to be available in a particular form—clean and organized, usually as "tidy" analytical tables—and it produces results that need to be prepared for consumption by non-specialists. As stated above, it is the pivot of the arc, the core process, where the prior phases constitute pre-processing and following phases post-processing.

Zone *I* is also relatively easy to interpret: the functions in this group each involve understanding the relationship between the pipeline and the external world, the messy interface between the enterprise of data science and the variety of real world situations in which it operates and impacts. Again, this relationship has a dual character, based on the difference between input and output, reading from and writing to the world. Note also that there is an open-ended quality to the this relationship that is elided by circular representations—there is always a knowledge gap between one's understanding of a data product and its real effects on the world, effects which may in turn influence source data.

We may note in passing that *I* and *IV* can be contrasted in several ways—external vs internal, messy vs clean, exoteric vs esoteric, qualitative vs quantitative, existential vs essential, concrete vs abstract, periphery vs center, etc.

When it comes to zones *II* and *III*, the interpretation of results is less straightforward. This is because the reality of the kind of work performed in these areas is not as clear-cut as it is for *I* and *IV*. Both *II* and *III* exhibit an internal complexity not found in the others, and the two are less clearly separable from each other than they are from the other two. One reason for this complexity is that here pure and applied forms of knowledge intermingle in ways that defy easy description from an academic perspective.

For example, the work of "data wrangling," often considered essential to data science, spans the two domains and involves a complex mixture of specific technological know-how and general scientific principles. It turns out that the relationship between these kinds of knowledge is highly contested, as evidenced by the reception of Donoho's "50 Years of Data Science," which has been criticized for separating science from engineering and demoting the importance of the latter (Donoho 2017b). Regardless of the validity of this criticism, there is without doubt a long-standing conflict between computational data mining and statistical data analysis over what counts as valid forms of knowledge (Alvarado 2023), and this conflict emerges in the representation of zones *II* and *III* we find in our corpus.

We can take the conflict of interpretations over the status of technical knowledge in data science as a clue and use it to identify two broad dimensions that cross-cut the functions in zones *II* and *III*: technical know-how and abstract representation. We can reassign our labels to these dimensions, using the prime notation of *II′* and *III′* to indicate that they are transformation of our original themes.

Technical know-how *II′* involves expertise in developing and deploying software and hardware designed to handle data at scale, including high-performance computing, big data architectures (such as Hadoop and its descendants), and data-oriented programming languages and libraries. The topics associated with *II′* are highly specific and change rapidly relative to other forms of knowledge, and so are often omitted from, or under-represented in, academic curricula, even though to many they are the *sine qua non* of data science.

Abstract representation *III′*, on the other hand, involves expertise in areas ranging from how data are to be modeled for capture and analysis to how the results of analyses are to be presented to non-expert decision-makers. These areas of knowledge strive for formal generality over the long run; they are often expressed as grammars or design languages, frequently with visual modes (such as entity-relationship models and unified modeling language UML). They also include other forms of visualization, such as the plots developed for exploratory data analysis, such as box plots, and those used to represent statistical facts and analytical results in dashboards and infographics.

**The Four Areas, Plus One**

We are now ready to define and name the areas of data science expertise that emerge from an analysis of the pipeline considered as an arc. In each case, we want to identify the common context shared by the paired activities in each zone as well as the tension that exists between them by virtue of their occupying opposite sides of the pipeline. In many cases, although we can identify a shared theme in each zone's work, the reality is that practitioners do not always interact or share disciplinary homes. One of the benefits of this model will be to identify these points of synergy and to identify new disciplinary boundaries.

**Area I: Value**

The area of value is defined by the relationship of data science to the world from which it draws data and into which it inserts data products. More broadly, it concerns the primary motivations of data science—why do we practice data science in the first place? It combines the traditional discipline of ethics with the professional activities of business planning, policy making, developing motivations for scientific research, and other activities that have a direct impact on people and the planet. This is the area where we determine what we do versus what we do not do, in order to maximize societal and environmental benefit and minimize harm. It is also the area that looks inward to the other data science areas and provides guidance on such issues as algorithmic bias or open science. Common activities include the forming of value propositions that initiate data science projects, research into how data is created and used "in the wild," understanding the ethics of data acquisition, manipulation, communication, and sharing, and the application of data products in the world.

**Area II´: Design**

The area of design is defined by the relationship between human and machine forms of representation. This relationship is bidirectional: human-generated data flowing into the pipeline must be represented for machine consumption (H2M, or $H \rightarrow M$), while analytically transformed data going out must be represented for human consumption (M2H, or $M \rightarrow H$). This area therefore includes expertise in human-machine interaction as it appears at the points of both consuming data and producing data products. Activities here include the representation and communication of captured data for the work of analytics, e.g. in database modeling, the curation of data, and of complex data and analytical results to humans to drive decision-making and influence behavior. It also includes the making of things, with purpose (i.e. to solve problems) and intent (meaning, concision, focus). A key part of the area is the broad practice of what is often called visualization, the translation of complex quantitative information into visual (and other sensory) forms that non-experts can understand. In slightly more technical terms, the area of design focuses on what Zuboff called "informating," the process by which the world is represented for computation and analytics, and also by which analytical models and results are represented to the world (Zuboff 1995). These two processes often produce competing representations—a private one *of* the world for the data scientist, and a public one *for* the world of the results of analytics. One task of this area is to reconcile these two representations.

**Area III´: Systems**

The area of systems is defined by the technological infrastructure that is common to the pipeline but concentrated in the activities of wrangling data, deploying data products, and building out systems to support these activities at scale. This area includes expertise in infrastructure systems and architectures to support working with big data—big in terms of volume, velocity, and variety—and building high performance systems in both development and production environments. It includes the broad areas of hardware and software as such—computer technology as opposed to computer science. Key activities include developing cloud resources, building performant pipelines to ingest and aggregate data, developing networks of resilient distributed data, and writing and using software to accomplish tasks. This area is often referred to as "data engineering" or "machine learning engineering," which, according to Owen, "is most of what Data Science is and Statistics is not" (Owen 2015).

**Area IV: Analytics**

The area of analytics is defined by the practice of mathematical modeling based on data. This area includes what many consider to be the essence of data science, the combination of statistical methods with machine learning, along with information theory, optimization, network analysis, complexity theory, simulations, and other rigorous quantitative methods from a variety of fields. Although unified by a broad commitment to advanced mathematical models and computational algorithms, in reality this is a heterogeneous collection of competing schools and methods. Tensions include inference vs prediction, parametric vs non-parametric (kernel-based) methods, frequentist vs Bayesian statistics, analytic vs algorithmic solutions (including simulations), etc. Key activities include clustering, pattern recognition, regression, rule mining, feature engineering, model selection, performance evaluation, and a host of other activities. Although currently dominated by statistical methods, this area also includes the rule-based methods that dominated the field of artificial intelligence before the more recent successes of statistical learning and deep learning.

**Area V: Practice**

The preceding four areas each represent areas of foundational knowledge, forms of expertise that can be taught as more or less separate subjects. In practice, however, these areas represent the interlocking parts of a division of labor that are integrated in the pipeline. This area consists of actual activities that brings people together to combine expertise from each of the four areas. It is characterized by data science teams working together and with external parties to develop solutions and projects that are responsible, authentic, efficient, and effective. Practice is also where the core areas of data science come into contact with a broad spectrum of domain knowledge and real world problems. The following diagram shows the central, integrative role played by practice:

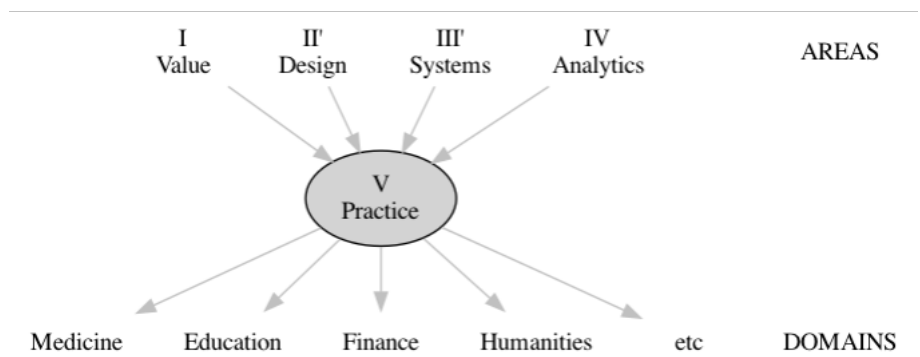

Figure 3. The Integrative Role of Practice

**Two Principal Components**

Is there a way to understand how the four primary areas are related to each other, beyond their being composed of functions from the same pipeline? Put another way, does the pipeline-as-arc model exhibit any structural features that will help us conceptualize the broader space of data science? Two such features stand out: (1) the opposition between concrete and abstract forms of representation, and (2) between human and machine processing.

Regarding the concrete and the abstract, it's clear that the arc model has a metric quality to it: as one moves toward the pivot point of analysis, one moves away from the concrete messiness of reality as experienced to the "tidy" and abstract world of mathematics; similarly, as one moves from the pivot back to the world, there is a requirement to convert esoteric results into more humanly intelligible forms, often through a process of concretization; visualizations succeed by employing concrete metaphors that flesh out mathematical ideas that are notoriously detached from the imagination—no one can imagine, for example, n-dimensional spaces beyond a handful of dimensions. The arc describes a dialectic of abstraction and concretization that defines the ebb and flow and data science work.

|  | Concrete | Abstract |
|---|---|---|
| Human | $I$<br>VALUE | $II'$<br>DESIGN |
| Machine | $III'$<br>SYSTEMS | $IV$<br>ANALYTICS |

Table 1. The Four Areas in Two Dimensions

The dimension of human and machine processing exhibits a similar duality, that between the conversion of information from humanly accessible forms, such as given by data acquired by instruments, into machine readable and processible forms, and the reverse. The process of moving from human to machine representations is a large part of what data capture, modeling, and wrangling is all about, while the process of converting the results of machine learning, broadly conceived, into humanly actionable form is what visualization and productization are all about. The reality of this dualism is captured by the concept of human-computer interaction (HCI), an established field that is applicable to both sides of the arc.

How do the four fundamental areas map onto these two dimensions? We can define each area as a combination of one pole from each duality; the four areas result from all possible permutations of the two dimensions. This produces the following high level characterizations of each area: (1) Value is concerned with concrete humanity, (2) Design with abstract humanity, (3) Analytics with abstract machinery, and (4) Systems is concerned with concrete machinery. All of these make intuitive sense, with the exception of Design. This is consistent, however, with the fact that the area of Design emerges from this analysis as an undervalued and not well understood area of expertise, even though Yau emphasized it early on (Yau 2009b). Indeed, one of the consequences of this analysis is to train our attention on this area of knowledge and to develop it further.

One exciting interpretation of the two dimensions defined here is that they correspond to two principal components that undergird the general field of data science. As components, these axes define two orthogonal dimensions within which all the specific topics of data science may, in principle, be plotted. The reality behind these axes may be that they represent cognitive styles associated with the division of labor implied by the data science pipeline.

**PC1: HUMAN VERSUS MACHINE**

The human-machine axis accounts for the most variance in the field. This seems evident from the fact that Conway's Venn diagram model of data science represents only the machine side of our model, with practice replaced by "substantive expertise" (Conway 2010b). The human side—Value and Design—is left out, or short-changed by being lumped in with domain knowledge. The very fact that the human side has to be explained and added to the model suggests strongly that it defines a pole at some distance from the areas of knowledge described in Conway's model. The human pole refers to humanity understood as situated in their historical, social, and cultural milieu. It is synonymous with *human experience*. The machine pole refers to the technoscientific apparatus of formal, quantitative reasoning that operates on representations of the human and the world. In the context of data science, it is more or less synonymous with *machine intelligence*, broadly conceived to include machine learning but also other modes of analysis on the spectrum of prediction and inference. Given these poles, the human-machine axis represents the opposition between humanistic disciplines that seek to understand human experience as such, and the formal sciences that employ machine intelligence, broadly conceived, to interpret that experience as represented and aggregated in the form of data.

**PC2: CONCRETE VERSUS ABSTRACT**

The abstract-concrete axis accounts for the difference between two forms of knowledge, roughly between direct experience and the indirect representation of that experience enabled through data. Both the realm of Value and Systems involve immersion in the messy details of lived experience— and direct acquaintance with the devils in those details. This is the messy world of hacks and ironies. The realms of Design and Analysis, on the other hand, are founded on abstract representations that strive for clear and distinct purity, and which allow for deductive reasoning to succeed at the cost of simplifying assumptions and reduced representations. This is the orderly world of models. The concrete pole refers to situated knowledge, knowledge as understood by hackers and makers, but also ethnographers who seek to maximize thick description in their work. It represents *concrete materiality*. The abstract pole refers to formal knowledge, knowledge in the form of mathematical symbolism, deductive proofs, and algorithmic patterns. It is *abstract form*. Given these poles, the concrete-abstract axis is roughly the opposition between applied and pure forms of knowledge, between those that embrace materiality and those that seek purity of form.

**Final Representation**

The result of the preceding may be represented by the following graphic.

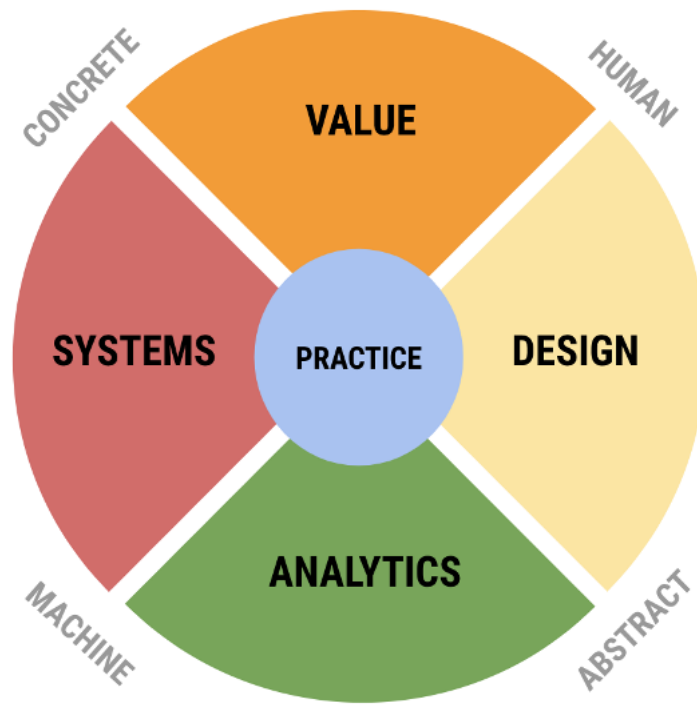

The 4+1 Model of Data Science

This visualization represents data science as composed of specific and complementary forms of knowledge. The vertical axis defines the dominant polarity between analysis—the *how* of data science, often identified entirely with it, contrasted with the *why* of data science, from which data science derives its meaning and value as a profession. The horizontal access defines the polarity of methods that are often obscured in academic definitions of data science—the supporting practices that make the Analytics component work in the first place.

**Concluding Remarks**

The point of the 4+1 model, abstract as it is, is to provide a practical template for strategically planning the various elements of a school of data science. To serve as an effective template, a model must be general. But generality if often purchased at the cost of intuitive understanding. The following caveats may help make sense of the model when considering its usefulness when applied to various concrete activities.

**The model describes areas of academic expertise, not objective reality**. It is a map of a division of labor writ large. Although each of the areas has clear connections to the others, the question to ask when deciding where an activity belongs is: *who would be an expert at doing it?* The realms help refine this question: the analytics area, for example, contains people who are good at working with abstract machinery. The four areas have the virtue of isolating intuitively correct communities of expertise. For example, people who are great at data product design may not know the esoteric depths of machine learning, and that adepts at machine learning are not usually experts in understanding human society and normative culture.

**Each area in the model contains a collection of subfields that need to be teased out**. Some areas will have more subfields than others. Although some areas may be smaller than others in terms of number of experts (faculty) and courses, each area has a major impact on the overall practice of data science and the quality of an academic program's activities. In addition, these subfields are in an important sense "more real" than the categories. We can imagine them forming a dense network in which the areas define communities with centroids, and which are more interconnected than the clean-cut image of the model implies.

**The principal components abstract/concrete and human/machine are meant to help imagine the kinds of activities that belong in each area**, through their connotations when combined to form the four bigrams—concrete human, abstract human, concrete machine, and abstract machine. For example, the area of value as the realm of the "concrete human" (or perhaps "concrete humanity") is meant to connote what the Spanish philosopher Unamuno called the world of "flesh and bone" within which we live and die, that is, where things matter. On the other hand, analytics as the realm of the "abstract machine" is meant to connote the platonic world of mathematical reasoning which, since Euclid, has been characterized by rigorous, abstract, deductive reasoning that has literally been described as an abstract machine (see Alan Turing).

**Appendix A: About the Sources**

The primary sources on which the conclusions of this essay are based comprise a variety of documents, from technical journals to blog posts to internal reports. They come from a range of viewpoints, from data analysis and statistics to data mining and data science *per se*. For the purposes of the essay, we select a more or less representative subset across these axes of variation. With respect to representativeness, in some cases a document was chosen for its influence, in others, such as the post by Dataman, because it is considered more or less typical of a common genre.

The documents chosen are listed below in chronological order, beginning with Tukey's seminal essay on data analysis and ending with contempory explainers. Included also are the definitions of the CRISP-DM and KDD processes which are the most developed pipeline models.

Each source entry below contains a short description of the source and its context, and then a list of the phases cited by the authors as fundamental to data processing. These phases are also mapped onto the standard sequence described in the main part of this essay, listed here for convenience.

1. Understand
2. Plan
3. Collect
4. Store
5. Clean
6. Explore
7. Prepare
8. Model
9. Interpret
10. Communicate
11. Deploy
12. Reflect

Mappings are indicated by an arrow pointing to the subset of terms from the standard sequence, e.g. … → [*Explore*] These mappings are also aggregated into a composite pipeline and displayed the table below; each model row is referenced by its key as defined in the entries.

Note that in most cases these phases are explicitly described as a process and often as a pipeline. When they are not, the implication is strong. In some cases, the process is likened to a cycle, emphasizing the connection between the endpoints of the pipeline, which is also emphasized by the 4+1 model.

A final feature added to each entry is a two-value indicator of bias—statistics and data mining. This is meant to capture the intellectual origin of the model, given that statistics and data mining define the poles of one of main axes of variance that defines the field of data science. This difference roughly corresponds to the "two cultures" described by Breiman Breiman (2001).

**LIST OF SOURCES**

Tukey on Data Analysis

Key: `Tukey`
Year: 1962
Source: Tukey (1962) URL
Bias: Statistics

In this classic essay, Tukey introduces the concept of data analysis, which he distinguishes from mathematical statistics and likens to an empirical science. He defines data analysis as an empirical process with phases including "… procedures for **analyzing** data, techniques for **interpreting** the results of such procedures, ways of **planning** the **gathering** of data to make its analysis easier, more precise or more accurate, and all the machinery and results of (mathematical) statistics which apply to analyzing data" (p. 2). Unpacking this statement yields a four phase model.

1. **Planning**: This phase includes "ways of planning the the gather of data to make its analysis easier." → $[Plan]$

2. **Gathering**: The gathering of data, either through creation or by acquisition of "data already obtained" (p. 40). Includes also the shaping of data "to make its analysis easier," which corresponds to our concept of Preparation. → $[Collect, Prepare]$

3. **Analyzing**: This is where data are analyzed with "all the machinery and results of (mathematical) statistics." → $[Explore, Model]$

4. **Interpreting**: "techniques for interpreting the results of" analysis. → $[Interpret]$

Fayyad on KDD

Key: `KDD`
Year: 1996
Source: Fayyad et al. (1996) URL→
Bias: Data Mining

KDD, or Knowledge Discovery in Databases, emerged in the late 1980s as both datasets and the computational resources to work with them became abundant. These resources included commercial databases and personal computers. In many ways the most adjacent field to contemoporary data science, this approach is unabashedly dedicated to finding patterns in data prior to developing a probabilistic model to justify their use. Fayyad's essay identifies five steps (Fayyad et al. 1996: 84). He emphasizes the highly iterative and cyclical nature of the process, arguing that it "may contain loops between any two steps." Another significant aspect of this conception of the pipeline is the role of exploration in the analytical phase: "Data Mining is a step in the KDD process consisting of applying data analysis and discovery algorithms that, under acceptable computational efficiency limitations, produce a particular enumeration of patterns over the data …." (p. 83)

1. **Selection**: Creating a target data set, or focusing on a subset of variables or data samples, on which discovery is to be performed. → $[Collect]$

2. **Pre-processing**: Cleaning and pre processing the data in order to obtain consistent data. → $[Clean]$

3. **Transformation**: Transformation of the data using dimensionality reduction and other methods. → [*Prepare*]
4. **Data Mining**: Searching for patterns of interest in a particular representational form, depending on the DM objective (usually, prediction). → [*Model*]
5. **Interpretation/Evaluation**: Interpretation and evaluation of the mined patterns. → [*Interpret*]

Azevedo on SEMMA

Key: `SEMMA`
Year: 1996
Source: Azevedo and Santos (2008)
Bias: Statistics

The SEMMA model was developed the by SAS institute in 1996 as part of the documentation for their product, SAS Enterprise Miner. Even so, the model is referenced outside of this context, often as a comparison to KDD and CRISP-DM. Its bias towards statististics is evident in the first step.

1. **Sample**: Sampling the data by extracting a portion of a large data set big enough to contain the significant information, yet small enough to manipulate quickly. → [*Collect*]
2. **Explore**: Exploration of the data by searching for unanticipated trends and anomalies in order to gain understanding and ideas → [*Explore*]
3. **Modify**: Modification of the data by creating, selecting, and transforming the variables to focus the model selection process → [*Prepare*]
4. **Model**: Modeling the data by allowing the software to search automatically for a combination of data that reliably predicts a desired outcome. → [*Model*]
5. **Assess**: Assessing the data by evaluating the usefulness and reliability of the findings from the DM process and estimate how well it performs. → [*Interpret*]

Hayashi on Data Science

Key: `Hayashi`
Year: 1998
Source: Hayashi et al. (1998) URL→
Bias: Statistics

The Japanese statistician Chikio Hayashi adopted the term "data science" in the early 1990s to define a field that did not succumb to what he saw to be the errors of both statistics and data analysis. He argued that mathematical statistics had become too attached to problems of inference and removed from reality, while data analysis had lost interest in understanding the meaning of the data it deals with. His definition of data science is decidedly processual: "Data Science consists of three phases: design for data, collection of data and analysis on data. It is important that the three phases are treated with the concept of unification based on the fundamental philosophy of science …. In these phases the methods which are fitted for the object and are valid, must be studied with a good perspective." (p. 41) Similar to KDD and CRISM-PM, Hayashi envisioned this process as a spiral, oscillating between poles if what he called "diversification" and "simplification." Note also that each of these terms, as described, comprises more than on of the standard sequence phases.

1. **Design**: Surveys and experiments are developed to capture data from "multifarious phenomena." → [$Understand, Plan$]
2. **Collection**: Phenomena are expressed as multidimensional or time-series data; properties of the data are made clear. At this stage, data are too complicated to draw clear conclusions. (Representation) → [$Collect, Explore, Prepare$]
3. **Analysis**: By methods of classification, multidimensional data analysis, and statistics, data structure is revealed. Simplification and conceptualization. Also yields understanding of deviations of the model, which begins the cycle anew. (Revelation) → [$Model, Interpet$]

Wirth and Hipp on CRISP-DM

Key: `CRISPDM`
Year: 1999
Source: Wirth and Hipp (1999b) URL→
Bias: Data Mining

By the late 1990s, the practice of data mining had become widespread in industry and globally. In 1999 the Cross Industry Standard Process for Data Mining (CRISP-DM) was developed in Europe as a comprehensive and general model to support the use of data mining in a broad range of sectors in a principled manner. Designed to work within a project management framework, this model is by far the most developed, and it continues to influence the field of data science to this day. Like KDD before it, the model emphasizes the cyclic and recursive nature of the process, and this perspective is reflected in a circular diagram that often accompanies its presentation. The steps below are based on the summary presented in Wirth and Hipp's essay.

1. **Business Understanding**: Understanding project objectives and requirements from a business perspective. Includes the development of a plan. → [$Understand, Plan$]
2. **Data Understanding**: The initial data collection and activities to get familiar with the data, e.g. to identify data quality problems, to discover first insights into the data, or to detect interesting subsets to form hypotheses for hidden information. This is really two phases—**Collection** and **Exploration**—which are combined because of their close, iterative relationship. → [$Collect, Explore$]
3. **Data Preparation**: Construction of the final dataset for analytical use. Tasks include table, record, an attribute selection, data cleaning, construction of new attributes, and transformation of data for modeling tools. → [$Clean, Prepare$]
4. **Modeling**: Modeling techniques are selected and applied, parameters calibrated. Modeling techniques include a broad range of unsupervised and supervised methods. As with KDD, there is an emphasis on pattern discovery, which has the effect of promoted methods that other models place squarely in the Explore phase of the standard sequence. → [$Model$]
5. **Evaluation**: Evaluation of model performance by both intrinsic and extrinsic measures. Regarding the latter, a key objective is to determine if an important business issue has not been sufficiently considered. → [$Interpret$]
6. **Deployment**: The knowledge gained by the model is presented in a way that the customer can use it. This may be something as simple as a report or as complex as a repeatable data mining process. In many cases the user, not the data analyst, will carry out the deployment. → [*Deploy*]

Mason and Wiggins on OSEMI

Key: `OSEMI`
Year: 2010
Source: Mason and Wiggins (2010a) URL→
Bias: Data Mining

After the phrase "data science" went viral (circa 2009), there were many efforts to make sense of the idea. In 2010 Drew Conway posted his Venn diagram of data science (Conway 2010a). The same year, another influential model, based explicitly on the pipeline, came from Mason and Wiggins in a blog post hosted at O'Reilly's Tech Radar site. In contrast to previous models rooted in statistics, this model assumes that data are abundant and available, such as data scrapable from the Web.

1.  **Obtain**: Gather data from relevant sources through APIs, web scraping, etc. → [*Collect*]

2.  **Scrub**: Clean data and convert data to machine readable formats. Clearning includes handling missing data, inconsistent labels, or awkward formatting; stripping extraneous characters; normalizing values, etc. → [*Clean, Prepare*]

3.  **Explore**: Find significant patterns and trends using statistical and data analytic methods, such as visualizing, clustering. Also includes transformations of the for more effective analysis, such as dimensionality reduction. → [*Explore*]

4.  **Model**: Construct methods to predict and forecast. These methods include those of inferential statistics and predictive machine learning. → [*Model*]

5.  **Interpret**: Making sense of the results as well as evaluating the performance of models. May involve domain experts. Also includes methods such as regularization that make models interpretable to those who use them, e.g. scientists or business people. → [*Interpret*]

Ojeda, et al. on Data Science

Key: `Ojeda+`
Year: 2014
Source: Ojeda et al. (2014) URL→
Bias: Data Mining

By 2014, data science had become a widespread practice in industry and the academic, and explanations of its nature became the subject of many books. This text is one of a genre that presents the field as a process, perhaps due to the influence of the CRISP-DM and OSEMI models, and uses the expression pipeline throughout. Note that the model defined in this book is not presented here as canonical. It suffers from various inconsistences, such as the labeling of steps in the text representation of the pipeline versus those on diagrams. It is included to demonstrate the pervasiveness of the model.

1.  **Acquisition**: Acquire the data from relational databases, NoSQL and document stores, web scraping, distributed databases (e.g. HDFS on a Hadoop platform), RESTful APIs, flat files, etc. Consistent with the other data mining models, the emphasis here is on working with available data, not generating it. → [*Collect*]

2. **Exploration and understanding**: Understand the data and how it was collected or produced; this often requires significant exploration. Note that this step does *not* correspond to exploration in the sense of exploratory data analysis (EDA). Rather, it reflects the position of the data scientist as the receiver of someone else's data and the need to infer what would normally belong to the first step of the standard squence $Understand$. → [$Understand$]

3. **Munging, wrangling, and manipulation**: Convert the data into the form required for analysis. This includes a wide range of activities, such as those mentioned in previous models. However, it also conflates the standard phases $Clean$ and $Prepare$. → [$Clean, Prepare$]

4. **Analysis and modeling**: Apply statistical and machine learning methods, including clustering, categorization, and classification. One presumes that the standard step of $Explore$ is included here. → [$Explore, Model$]

5. **Communicating and operationalizing**: At the end of the pipeline, we need to give the data back in a compelling form and structure, sometimes to ourselves to inform the next iteration, and sometimes to a completely different audience. The data products produced can be a simple one-off report or a scalable web product that will be used interactively by millions. → [$Communicate, Deploy$]

Caffo, et al. on Data Science

Key: `Caffo+`
Year: 2015
Source: Caffo, Peng, and Leek (2015) URL→
Bias: Statistics

By 2015, many universities had begun offering degrees in data science, typically at the masters' level, with the intention of meeting the high demand for data scientists. Professors Caffo, Peng, and Leek's book was written to accompany a course in Exectutive Data Science, offered by Johns Hopkins University through *Coursera*. Their model is relatively high level, consisting of five phases, given the target audience of those in charge of data science teams. As with other models, this model emphasizes the iterative nature of each phase, both internally and between phases. And as with many statistics-oriented conceptions of data science, this model emphasizes the Understand phase and skips over the technical issues of storing and modeling the data.

1. **Question.**: Pose a research question and specify what is to be learned from data to answer it. The question determines the data to be obtained and the type of analysis to perform. Included determing the type of question, including descriptive, exploratory, inferential, causal, predictive, and mechanistic. An alternate approach here is *hypothesis generation*, which may be suitable when data already exist but a question is not well-developed. In this scenario, the data scientist may skip to the next step to determine the value of the data. Once a question is developed, then it may be necessary to acquire more data, and then go through the process. → [$Question, Collect$]

2. **Exploratory data analysis**: Explore the data to determine if the data are suitable for answering the question and if more data need to be collected. For example, determine if there are enough data and if it is missing key variables. In addition, develop a sketch of the solution. Include a freamework for challenging results and to develop robust evidence for answering your question. → [$Explore$]

3. **Formal modeling**: Identify the parameters to estimate based on the sketch. → [*Model*]
4. **Interpretation**: Determine if the modeling results align with the initial expectations during the Question phase and before the acquisition of data. Consider the totality of the evidence developed after attempting to fit different models, weighing the different pieces of evidence. → [*Interpret*]
5. **Communication**: Communicate findings to various audiences, either internal to the organization or external. Includes translating findings into action by virtue of effectively communicating results to decision-makers. → [*Communicate*]

Donaho on Data Science

Key: Donoho
Year: 2017
Source: Donoho (2017a) URL→
Bias: Statistics

As data science became viral in the 2010s, academic statisticians frequently expressed concern that they were "disconnected from the new (and vaguely defined) community of data scientists, who are completely identified with Big Data in the eyes of the media and policymakers" (Rodriguez 2012). "Aren't *We* Data Scientists?" asked Marie Davidian, then president of the American Statistical Association, in 2013 (Davidian 2013). In response to this growing sentiment, Donoho's essay reads as a manifesto for the reclaiming of data science by academic statistics. In it, he defines six divisions of Greater Data Science, each containing a set of subactivities that roughly map to the pipeline model described here.

It is important to note that Donoho's model is more abstract than a pipeline description and therefore not all of the divisions and subactivities directly map onto the sequence. **Data visualization and Presentation** defines a general practice, although from the description it clearly maps onto two phases, Explore and Communicate. **Computing with Data** refers to knowledge of programming languages for data analysis and data processing as well as knowledge of how to use cluster and cloud computing resources at scale. It also includes how to develop workflows which organize work. Clearly, this area belongs to no phase in particular but instead characterizes the broader context in which the data science pipeline operates. The identification of workflows, which are the focus of the Science about Data Science division, also suggests that Donoho is working at a higher level of abstraction than the other models, which places it alongside the of the current essay. The following phases are inferred from Donoho's descriptions.

1. **Gathering**: This includes both experimental design, modern data gathering techniques, and identification of existing data resources, from signal data to websites. → [*Plan, Collect*]
2. **Preparation**: Identification of anomalies and artifacts in the data and handling them by reformatting, recoding, grouping, smoothing, subsetting, etc. → [*Clean*]
3. **Exploration**: Application of EDA to sanity-check data and expose unexpected features. Includes data visualization, which Donoho separates out into a separate division and combines with visualization activities involved in interpretation and communication. → [*Explore*]
4. **Modern Databases**: Transform and restructure data as found in source files, such as CSV files and spreadsheets, and databases, into a forms more suitable for analysis. → [*Prepare*]

5. **Mathematical Representations**: Application of mathematical structures for to extract features from special kinds of data, including acoustic, image, sensor, and network data. For example, the application of the Fourier transform to acousting data or the wavelet transform to image data. → [$Prepare$]
6. **Data Modeling**: Appliction of methods from both traditional statistics and contemporary machine learning. → [$Model$]
7. **Presentation**: The creation of sophisticated graphics, dashboards, and visualizations to present conclusions to stakeholders. → [$Communicate$]
8. **Science about Data Science**: In the spirit of Tukey's "science of data analysis," this is the evaluation of what data scientists actually do and produce. Includes the identificatin and study of commonly occurring analytical and processing workflows. → [$Reflect$]

Géron on Machine Learning

Key: `Géron`
Year: 2017
Source: Géron (2017) URL→
Bias: Data Mining

Géron's text is a classic among practicing data scientists interested in using machine learning in a business setting, covering everything from regression to deep learning from a practical, code-centric perspective. Written with "minimal theory," the book demostrates the entrenched nature of the pipeline model, especially as it has been become a software development pattern hard-coded into both SciKit Learn and TensorFlow. This usage reflects the fact that within machine learning, "pipeline" has taken on a more specific meaning—"a sequence of data processing components"—than we are using here. These components are units software within a system, not the phases of labor associated with the work of the data scientist. Nevertheless, Géron's text describes a labor pipeline within which the software pipeline is embedded, the steps of an "end-to-end" classification project.

1. **Look at the big picture**: Frame the problem by defining the business objective and the specific goals of the model. This may include defining a specific performance measure, such as a loss function. Consider that the model is a a means to an end. → [$Understand$]
2. **Get the data**: This consists of setting up a compute workspace and downloading the data. This step also includes getting to know the data and preparing a test set. → [$Get, Prepare$]
3. **Discover and visualize the data** to gain insights: Go into more depth exploring the data, using EDA methods to investigate correlations and experiment with attribute combinations. → [$Explore$]
4. **Prepare the data** for Machine Learning algorithms: This step involves transforming and structuring the data in forms suitable for the algorithms that with fit the data to a model. This includes imputing missing data, handling non-numeric data, feature scaling, etc. This step contains its own pipeline. → [$Clean, Prepare$]
5. **Select a model and train it**: Apply models deemed appropriate to the data and compare results. Apply evaluation methods such as cross-validation to compare model results. → [$Model$]

6. **Fine-tune your model**: Once the list of candidate models is shortened, fine-tune their parameters by using various seach methods, e.g. grid, randomized, or ensemble. Also includes evaluating the models on test sets. → [*Model*]

7. **Present your solution**: : This step includes presenting to stakeholders what was learned, what worked and what did not, what assumptions were made, and what the system's limitations are. It also includes documenting everything, creating user-friendly presentations with clear visualizations and easy-to-remember statements. Géron refers to this as the "prelaunch phase," presumably because the component must be approved to go on to the next phase. → [*Communicate*]

8. **Launch, monitor, and maintain your system**: This step includes converting your model into a production-ready component that can become a functioning piece of the overall pipeline. This may mean creating a web service. → [*Deploy*]

Das on Data Science

Key: `Das`
Year: 2019
Source: Das (2019) URL→
Bias: Data Mining

This essay belongs to a genre of self-publication that attempts to explain concepts in data science to the public. It is typical of platforms like *Medium* and what used to be called the blogosphere. It is included here to represent the commonplace nature of the pipeline as a rhetorical device for explaining data science. Here, the pipeline is called a "life-cycle," although the term pipeline is used as well. The cyclical nature of the process is emphasized by including the first step as last step of the process. [Note that this essay was removed from the web by the author; a link to Internet Archive URL is included for completeness.]

1. **Business Understanding**: Understand the problem you are trying to solve and how data can be used to support a solution or decision. Identify central objectives and variables that need to be predicted. (Here the author implies that methods such as regression and clustering are objectives.) → [*Understand*]

2. **Data Mining**: Gathering the data from various sources. This may invovle extracting data from a legacy database or webscraping. The author correctly notes that this step should not be lumped together with cleaning. → [*Collect*]

3. **Data Cleaning**: This step includes cleaning and preparing the data, also know as "data janitor work." This step takes most of the data scientist's time because there are so many reasons that data may need cleaning. Also includes handling missing data. → [*Clean*]

4. **Data Exploration**: This is the brainstorming phase of data analysis, where one discovers patterns and biases in the data. Involves using the basic tools of EDA but also creating interactive visualizations to allow drilling down into specific points, e.g. to explore the story behind outliers. Here also one begins to form hypotheses about the data to be developed. → [*Explore*]

5. **Feature Engineering**: This is the process of using domain knowledge to transform the data into informative features that represent the business problem. This stage directly influences the accuracy of the predictive model constructed in the next stage. Methods include feature selection (i.e. dimensionality reduction) and constructing new features that will aid in the modeling process. → [*Prepare*]

6. **Predictive Modeling**: The application of machine learning methods to the data. Includes training several models and evaluating their performance, as well as applying statistical methods and tests to ensure that the outcomes from the models make sense and are significant. Based on the questions developed in the business understanding stage, this is where a model is selected. Model selection will depend on the size, type and quality of the data, availability of computational resources, and the type of output required. → [*Model*]

7. **Data Visualization**: This step combines expertise from the fields of communication, psychology, statistics, and art, with an ultimate goal of communicating the insightw from the model in a simple yet effective and visually pleasing way. → [*Communicate*]

Dataman on Data Science

Key: `Dataman`  
Year: 2020  
Source: Dataman (2020) URL→  
Bias: Data Mining

Another example of a self-published explainer essay, this one describes the data science "modeling process" and aligns it with six consultative roles. The other defines eight steps to the process. Curiously, althhough this pipeline focuses on the details of training models, it does not include training the model itself as a step.

1. **Set the objectives**: This step includes defining the goals of the model as well as its scope and risk factors. These will determine what data to collect, and whether the cost to collect the data can be justified by the impact of the model. → [*Understand*]

2. **Communicate with key stakeholders**: This step involves ongoing aligning expected outcomes with key stakeholders. This step is unique among the pipelines by being place so early in the process. We associate it with the *Understand* phase because it essentially broads the group for whom understanding matters. → [*Plan*]

3. **Collect the necessary data for exploratory data analysis (EDA)**: This step combines the *Collect*, *Clean*, and *Explore* phases. Involves the iterative "curation" of data need to conduct EDA. → [*Collect, Clean, Explore*]

4. **Determine the functional form of the model**: In this step, the specific from of the model is defined, including the definition and characterization of the target variable. This step involves model selection and would in practice be closely associated with the next. → [*Prepare, Model*]

5. **Split the data into training and validation** This step is concerned with model validation and avoiding overfitting. Data are divided into training and test datasets. It is assumed that test data were separated out prior to the preceding step. Presumably this step includes fitting the models, but this is not explicit. → [*Prepare, Model*]

6. **Assess the model performance**: This step includes determining the stability of a model over time (generalizability), focusing on the overall fit of the model, the significance of each predictor, and the relationship between the target variable and each predictor. Includes measures such as lift. Clearly this step follows the process of fitting and tuning models. → [$Model$]

7. **Deploy the model for real-time prediction**: The deployment of machine learning models into production, e.g. via batch prediction as a webservice. → [$Deploy$]

8. **Re-build the model**: This step involves revisiting the pipeline as models lose their predictability due to a variety of causes. Effectively, this step asserts the cyclic and interative nature of the pipeline and therefore belongs to no step in particular.

Porter on Data Science

Key: `Porter`
Year: 2020
Source: Porter (2020)
Bias: Statistics

Michael Porter is an Associate Professor of Data Science and Systems Engineering at the UVA. This essay is an internal report (available on request) on the field of data science from the perspective of curricular planning. Porter argues that Data Science includes seven areas, each of which can be viewed as a science, i.e. as requiring specific expertise. Like Donoho's essay (and the current), the model presented is more abstract than a pipeline model and includes areas that cross-cut steps in the Primary Sequence. Nevertheless, it retains a sequential structure consistent with the general pattern.

1. **Data Collection and Acquisition**: The science of "how" and "when" data is collected, and includes all methods of data acquisition from its production through designed experiments to its consumption from external sources, e.g. databases and APIs. → [$Collect$]

2. **Data Storage and Representation**: The science of "how" and "when" data is collected, including data modeling and storing data in databases and files of various formats. Also includes transforming data into "tidy" format. → [$Store$]

3. **Data Manipulation and Transformation**: The science of preparing data for analysis, including wrangling, cleaning, and importing data from databases (after they have been stored in the previous step). → [$Clean, Prepare$]

4. **Computing with Data**: The science of computing for data analysis with a focus on algorithm design and performance evaluation. → [$Model$]

5. **Data Analytics**: The science of machine learning, broadly conceived to include methods ranging from geneative modeling (either frequentist or Bayesian) and inference to predictive modeling and optimization. Notably, this step also includes EDA and feature engineering. → [$Explore, Prepare, Model$]

6. **Summarizing and Communicating Data and Models**: The science of extracting and summarizing the information in data for human understanding. This area includes visualization in the context of both EDA and presentation of results to external stakeholders. It also includes the communication of model and data properties (such as bias) to guide interpretation of results. Here we map it to the latter, but note that this area includes at least two steps. In addition, we may may the work of summarization to interpretation. → [$Interpret, Communicate$]

7. **Practicing Data Science**: The science of the overall system of data science, including improving the data science spipeline, replicability of results, openness and transparency, project management, etc. → $[Reflect]$

8. **Disciplinary Data Science**: The science of applying data science to specific disciplines. This involves a consideration of how the pipeline operates in different contexts, including how domain knowledge informs each of the steps of the pipeline, from mode of data acquisition to model selection and analytic appraoch, to the interpretation and communication of results. Although placed at the end of the list, it properly belongs to the initial *Understand* step. → $[Understand]$

**SUMMARY TABLE**

| | understand | plan | collect | store | clean | explore | prepare | model | interpret | communicate | deploy | reflect |
|---|---|---|---|---|---|---|---|---|---|---|---|---|
| Tukey | | ■ | ■ | | | ■ | ■ | ■ | ■ | | | |
| KDD | | | ■ | | ■ | | ■ | ■ | ■ | | | |
| SEMMA | | | ■ | | | ■ | ■ | ■ | ■ | | | |
| Hayashi | ■ | ■ | ■ | ■ | | ■ | ■ | ■ | | | | |
| CRISPDM | ■ | ■ | ■ | ■ | ■ | ■ | ■ | ■ | ■ | ■ | ■ | |
| OSEMI | | | ■ | | ■ | ■ | ■ | ■ | ■ | | | |
| Ojeda+ | ■ | | ■ | ■ | ■ | | ■ | | | ■ | ■ | |
| Caffo+ | ■ | ■ | ■ | | | ■ | | ■ | ■ | ■ | | |
| Donoho | | ■ | ■ | | | ■ | ■ | ■ | ■ | | ■ | ■ |
| Géron | ■ | | ■ | | | ■ | ■ | ■ | ■ | | ■ | ■ |
| Das | ■ | | ■ | | | ■ | ■ | ■ | ■ | | ■ | |
| Dataman | ■ | | ■ | | | ■ | ■ | ■ | ■ | | | ■ |
| Porter | ■ | | ■ | ■ | ■ | | ■ | ■ | ■ | ■ | | ■ |

**Appendix B: Relation to AI**

The four areas of data science defined here are surprisingly analogous to the four approaches to artificial intelligence defined by Russel and Norvig in their classic textbook on the subject (Russell and Norvig 1995: 5). Their four part model was generated by combining the axes *thinking vs acting* (behavior) and *human vs rational* (mode) as follows:

| behavior | mode | systems |
|---|---|---|
| thinking | humanly | cognitive models, ontologies |
| thinking | rationally | logic, laws of thought |
| acting | humanly | Turing test, situated action |
| acting | rationally | agents, bots |

It is easy to see how the following analogies, between the axes of the 4+1 model of data science and their model of AI, make sense:

| DS | AI |
|---|---|
| abstract : concrete | thinking : action |
| human : machine | human : rational |

Or, expressed as a set of identities:

| DS | AI |
|---|---|
| abstract | thinking |
| concrete | action |
| human | human |
| rational | machine |

So, it makes sense to compare the first table above with this one for data science:

| level | focus | area | topic |
|---|---|---|---|
| abstract | human | design | ontologies, data models, visualizations |
| abstract | machine | analytics | math, logic, algorithms |
| concrete | human | value | ethics, research questions, value propositions |
| concrete | machine | systems | hardware, software, security |

By comparing last columns of each table we can see that the 4+1 model of data science and Russell and Norvig's model of artificial intelligence share the same general space. The difference is that the former defines kinds of *acquired knowledge*, whereas the latter concerns kinds of *built systems*.